\documentclass[10pt]{iopart}

\usepackage{cite}
\usepackage{color}
\usepackage{graphicx}
\usepackage{iopams}

\usepackage{tikz}

\begin{document}

\title{Frequency conversion between UV and telecom wavelengths in a lithium niobate waveguide for quantum communication with Yb$^+$ trapped ions}

\author{Sachin Kasture$^{1}$, Francesco Lenzini$^{1}$, Ben Haylock$^{1}$, Andreas Boes$^2$, Arnan Mitchell$^2$, Erik W. Streed$^{1,3}$ and Mirko Lobino$^{1,4}$}

\address{$^1$Centre for Quantum Dynamics, Griffith University, 170 Kessels Rd, Nathan QLD 4111, Australia}
\address{$^2$ARC Centre for Excellence CUDOS, School of Engineering, RMIT University, 124 La Trobe Street, Melbourne VIC 3000, Australia}
\address{$^3$Institute for Glycomics, Griffith University, Parklands Drive, Southport QLD 4215, Australia}
\address{$^4$Queensland Micro and Nanotechnology Centre, Griffith University, 170 Kessels Rd, Nathan QLD 4111, Australia}

\ead{m.lobino@griffith.edu.au}
\vspace{10pt}

\begin{abstract}
We study and demonstrate the frequency conversion of UV radiation, resonant with 369.5~nm transition in Yb$^+$ ions to the C-band wavelength 1580.3~nm and vice-versa using a reverse proton-exchanged waveguide in periodically poled lithium niobate. Our integrated device can interface trapped Yb$^+$ ions with telecom infrastructure for the realization of an Yb$^+$ based quantum repeater protocol and to efficiently distribute entanglement over long distances. We analyse the single photon frequency conversion efficiency from the 369.525~nm  to the telecom wavelength and its dependence on pump power, device length and temperature. The single-photon noise generated by spontaneous Raman scattering of the pump is also measured. From this analysis we estimate a single photon conversion efficiency of $\sim$9\% is achievable with our technology with almost complete suppression of the Raman noise.
\end{abstract}

%
\vspace{2pc}
\maketitle
%
\ioptwocol

\section{Introduction}
Quantum information science aims at harnessing unique quantum mechanical properties such as quantum superposition and entanglement to deliver machines capable of performing  specific computational tasks \cite{Blatt-Shor,Lloyd1073} exponentially faster than classical computers, and to enable secure quantum communications \cite{Zhang}. Different hardware implementations are currently investigated for the realization of real world quantum devices \cite{Ladd} with integrated photonics \cite{shadbolt2012generating, Li_NJP} and trapped ions \cite{Debnath-Monroe-2016, Kielpinski2015} being two of the leading approaches.
Trapped ions have the advantages of being a fully scalable approach, where deterministic multi-qubit gates with a fidelity $>$97\% have been demonstrated \cite{Ion_fidelity}. Photons are excellent candidates for transferring quantum information over long distances due to their speed, the possibility of traveling inside optical fibers and their weak interaction with the environment. 

The first quantum technology that has reached commercial markets is quantum key distribution (QKD) \cite{Zhang}, which enables the secure sharing of a common key between two parties for the encryption of a message. As the no-cloning theorem does not allow for loss compensation through amplification, intrinsic optical fiber propagation losses limit the operational range of current systems to $\sim$200~km~\cite{QKD200km}. One way around these limitations is to use a quantum repeater protocol \cite{Briegel_Repeater} where the transmission line is divided into smaller segments connected by nodes that can store quantum information. In this protocol  entangled particles are first shared and stored in adjacent nodes, before the entanglement sharing is extended between distant nodes through entanglement swapping \cite{Ent_Swap}. Once the particles encoding the entangled state are shared between the parties, the secure key can be transmitted via quantum teleportation \cite{teleport}.

While photons are the only viable choice available for sending quantum information across a long distance, trapped ions are the perfect candidate for the implementation of the quantum repeater nodes \cite{Repeater_ion}. Ions have a long coherence times ($\sim$50s) which makes them very good quantum memories, and because of their strong interaction they can deterministically perform entanglement swapping operations \cite{swap_ion}. In spite of these advantages, ions fastest cycling transitions usually emit photons in the ultraviolet (UV) region of the spectrum and hence are unsuitable for long distance communication using optical fibers. 

Our device is a reverse proton exchanged (RPE) waveguide~\cite{LenziniOE, Korkishko98} in periodically poled lithium niobate (LN) and is an alternative technology to what proposed in \cite{Rutz2016UV} with  potassium-titanyl-phosphate (KTP) where the conversion was from UV to 1311~nm. Our material combines the high $\chi^{(2)}$ (second-order optical non-linearity) of LN with strong modal confinement and low propagation loss for efficient sum-frequency-generation (SFG) and difference-frequency-generation (DFG). In addition the symmetric index profile of the RPE waveguide improves the overlap integral of the interacting modes and it gives a coupling efficiency of $\sim$86\% at telecom with single mode optical fibers. This type of waveguides have been used for frequency conversion of single photons emitted by a quantum dot~\cite{de2012quantum} and, more recently, a silicon on insulator waveguide has been used for the frequency conversion of single photons around the telecom band~\cite{Bell:16}.

In this work we demonstrate the frequency conversion of 369.525~nm radiation, corresponding to the $^2$P$_{1/2}$$\rightarrow$$^2$S$_{1/2}$ dipole transition of Yb$^+$  to the telecom wavelength and vice-versa via difference and sum frequency generation (DFG and SFG) with a strong pump at 482.3~nm in a nonlinear optical waveguide. Finally we measured the noise generated by the pump laser through Raman scattered single photons in the telecom band and studied the performance of our device for the frequency conversion of single photons.


\section{Waveguide design and experimental set-up}\label{setup}
\begin{figure}[t]
\centering
\includegraphics[width=0.49\textwidth]{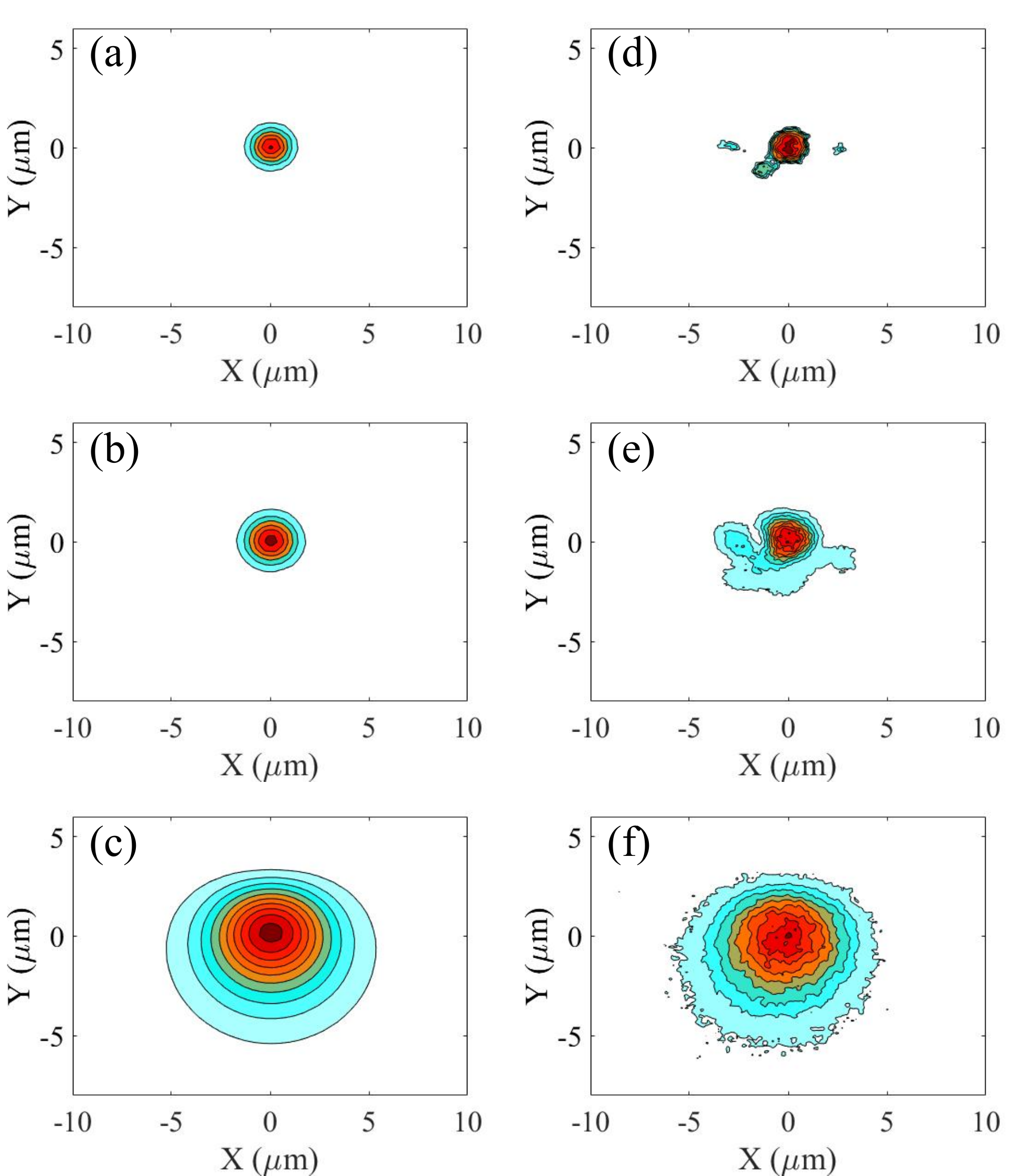}
\caption{(a)-(c) Simulated mode intensity profiles for 369, 482 and 1580~nm. (d)-(f) Measured intensity profiles for the same wavelengths.}
\label{modes}
\end{figure}

A 3~cm long and 10~$\mu$m wide waveguide was fabricated on a periodically poled Z-cut wafer of LN using the annealed reverse proton exchange technique \cite{LenziniOE,Korkishko98}. The waveguide was designed to be single mode at 1580~nm and the fabrication process consisted of four steps. First a top guiding layer of 1.9~$\mu$m depth is fabricated on the sample by proton exchange in pure benzoic acid at 170~$^\circ$C. Subsequently the sample is annealed in air at 328~$^\circ$C for 9~h and reverse exchanged in an eutectic melt of sodium nitrate, lithium nitrate and potassium nitrate \cite{RPE} at the same temperature for 15~h. Finally we performed another annealing step for 6~h at the same temperature. Figure \ref{modes} shows the simulated and measured mode intensity profiles at the three interacting wavelengths. These profiles accounts for an overlap integral between the modes 
\begin{eqnarray}\label{overlap}
\theta&=&\int\int^{+\infty}_{-\infty}E_{\omega_{1}}(x,y)E_{\omega_{2}}(x,y)E_{\omega_{3}}(x,y)dxdy \\
 &=&1.46\times10^5m^{-1}, \nonumber
\end{eqnarray} 
where $E_{\omega_i}(x,y)$ is the mode field profile normalized such as $\int\int^{+\infty}_{-\infty}\left|E_{\omega_{i}(x,y)}\right|^2dxdy$=1.

The substrate was periodically poled with a period of 7.11~$\mu$m corresponding to third order quasi-phase-matching (QPM) and resulting in an effective nonlinear coefficient d$_{\mbox{eff}}$=2d$_{33}$/3$\pi$ where d$_{33}$ is the bulk coefficient. This choice was necessary since at the 2.37~$\mu$m period required for first order QPM it is difficult to achieve a uniform poling because of non-uniform nucleation during the domain switching, spreading of domains below the electrodes, and domain-merging during the forward domain growth stage due to domain tip interaction. Propagation losses for our waveguide were measured to be 0.1~dB/cm at telecom, 0.7~dB/cm for the pump and 1.6~dB/cm for the UV.

Figure \ref{setup} shows the experimental set-up used for the frequency conversion measurements. The pump laser is a single spatial-mode temperature controlled Nichia diode tuned at  a 482.3~nm by a diffraction grating external cavity which also ensure a sub-MHz linewidth. The narrow linewidth is essential a coherent frequency conversion since the transition linewidth of $^{171}$Yb$^+$ is 19~MHz and the hyperfine splitting between the two qubit level is 12.6~GHz. The light is sent through a 35~dB optical isolator to reduce back reflections and improve the stability of the laser, followed by a half-wave plate and polarizing beam-splitter to control the pump power. The wavelength of the laser was continuously monitored on a wavemeter. For the SFG measurement, the pump laser is overlapped with the light coming from a tunable IR laser at a beam combiner and both beams are coupled into the waveguide by an aspheric lens with 0.68 NA and 3.1~mm focal length. The waveguide output is collected with another aspheric lens (NA=0.55 and focal length=4.5 mm) and sent through a series of filters to filter out the pump and the IR beams while the upconverted UV power is measured with a power meter. A similar scheme is used to measure the DFG of telecom radiation but the IR laser is replaced with a UV laser diode at 369.525~nm wavelength, near the Yb$^+$ transition, and at the output UV filters are changed to IR filters.

Both the input and the output lenses are mounted on a 3-axis micrometer stage assembly. The waveguide chip is mounted inside a PID-controlled oven with a temperature stability of 0.1~$^\circ$C on a 2-axis micrometer stage. We used a camera at the output to visualize the modes to ensure a high modal overlap of the fundamental mode for the pump and the UV where the waveguide is highly multimode. 

\begin{figure}[b]
\centering
\includegraphics[width=0.45\textwidth]{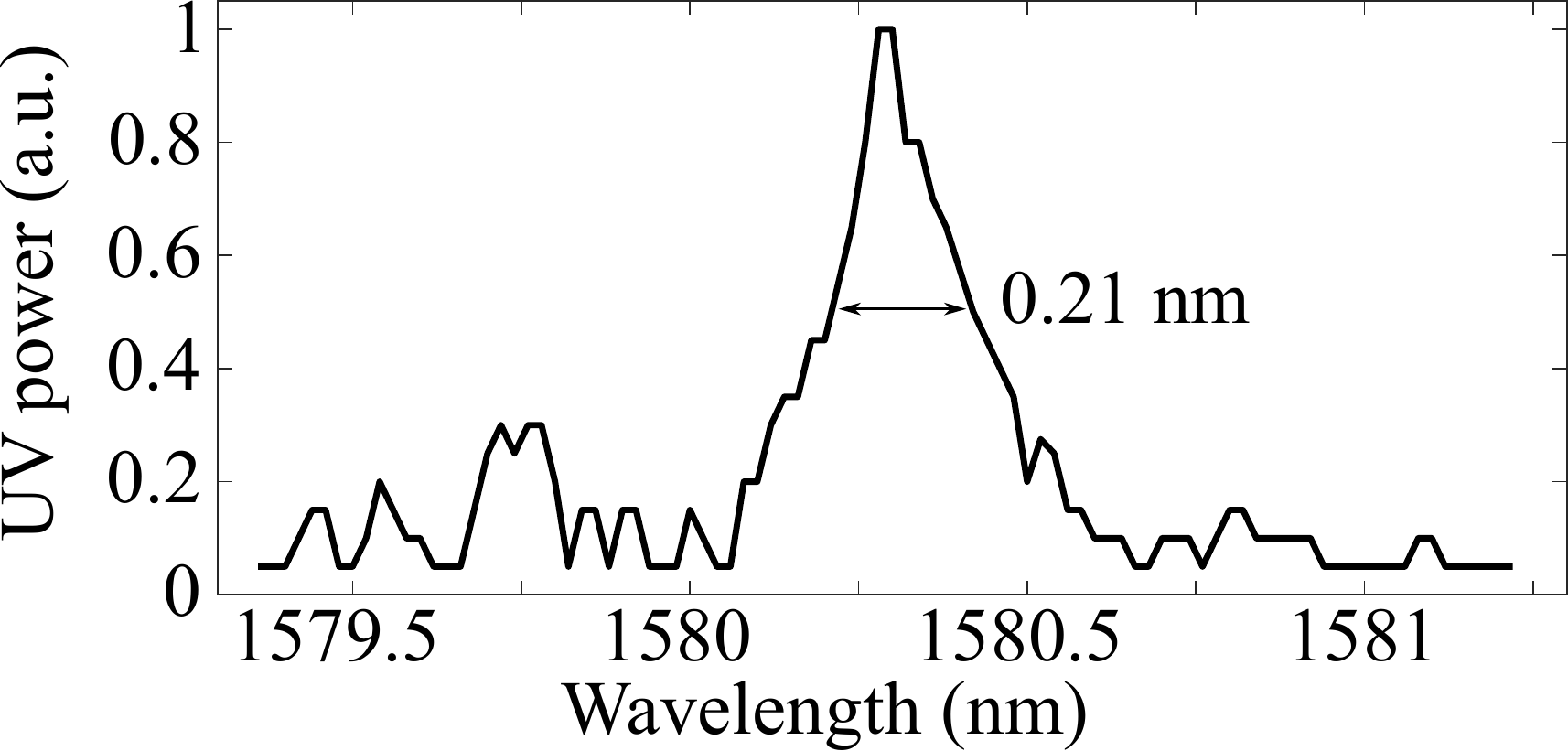}
\caption{Generated UV power as a function of the IR wavelength. From the FWHM of this curve we infer an interaction length of 4.9~mm for the SFG process.}
\label{sinc}
\end{figure}

To characterise the nonlinear performance of the waveguide we measured the generated UV power as a function of the IR laser wavelength (see Fig.~\ref{sinc}). The expected behaviour is that of a sinc function and the differences we see in our measurement are probably caused by temperature non-uniformity in the fabrication process resulting in inhomogeneity of the waveguide refractive index profile along its length. This non-uniformity was measured by reconstructing the refractive index profile on different parts of the wafer after proton exchange using the prism coupling technique. The measurements showed a parabolic variation of the refractive index across the device with a maximum change of 1\% at 635~nm wavelength. This variation is consistent with the temperature profile inside our reactor which is hotter in its centre. From the curve FWHM of 0.21~nm we estimate an interaction length of 4.9~mm for the SFG process which, together with the measured overlap integral of Eq.~\ref{overlap}, is consistent with the estimated conversion efficiency shown in Fig.~\ref{FreqConv}b. Coupling of the pump into higher order modes and non-uniformity in the poling pattern may also reduce the overall conversion efficiency in this device.

\begin{figure}
\centering
\includegraphics[width=0.45\textwidth]{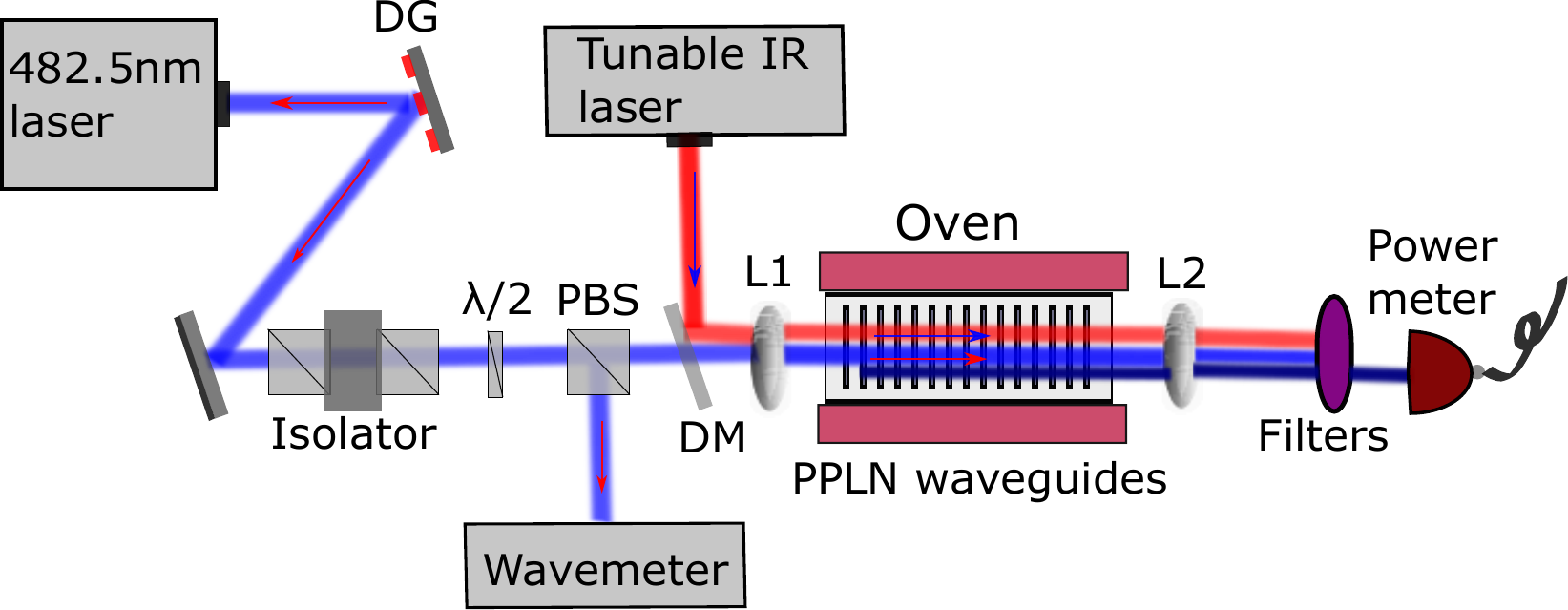}
\caption{Optical setup for sum frequency generation. The first order diffraction from the grating is used to provide feedback and tune the pump laser.  DG: diffraction grating, $\lambda$/2: half waveplate, PBS: polarizing beam splitter, DM: dichroic mirror, PPLN: periodically poled lithium niobate, L1 and L2: lenses. For the DFG, the tunable IR source is replaced by a UV laser and the power meter is replaced by a spectrum analyser.}
\label{set-up}
\end{figure}

\section{Frequency conversions}\label{results}
\begin{figure}[t]
\centering
\includegraphics[width=0.45\textwidth]{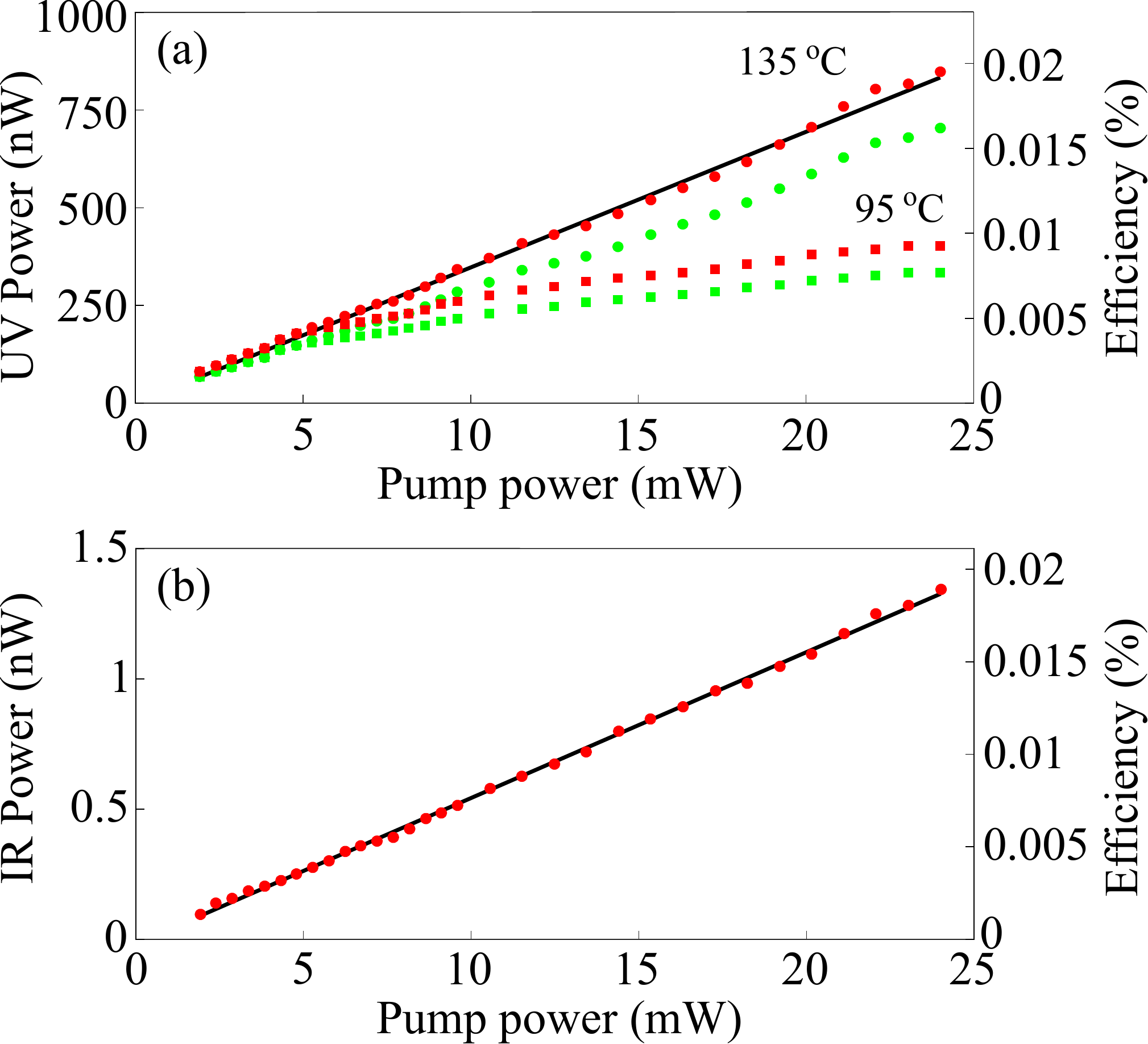}
\caption{(a) SFG output power and single photon conversion efficiency as a function of the pump power with (\textcolor{red}{$\bullet$} at 135$^\circ$C with theoretical prediction as solid line and \textcolor{red}{$\blacksquare$} at 95~$^\circ$C, internal conversion) and without (\textcolor{green}{$\bullet$} at 135$^\circ$C and \textcolor{green}{$\blacksquare$} at 95$^\circ$C, external conversion) Fresnel losses compensation. Coupled IR power is 1~mW. (b). DFG output power with internal single photon conversion efficiency (\textcolor{red}{$\bullet$}) and the theoretical fit (solid line) for a coupled UV power of 30~$\mu$W at a working temperature of 135~$^\circ$C.
}
\label{FreqConv}
\end{figure}

Figure \ref{FreqConv} shows the generation of UV from IR (SFG in (a)) and IR from UV (DFG in (b)) as a function of the pump power and the respective single photon conversion efficiencies are defined as
\begin{equation}
\label{eq:eff}
\eta=		\frac{P_{out}}{P_{in}}\times\frac{\lambda_{out}}{\lambda_{in}}\textrm{,}
\end{equation}
where P$_{out}$  is the converted power while P$_{in}$ is the input signal for SFG or DFG. In our experiment we were limited by the maximum pump power of  24~mW corresponding to 15.5~mW coupled into the waveguide. 

Frequency upconversion is shown in Fig.~\ref{FreqConv}a for 95$^\circ$C and 135$^\circ$C. At 95$^\circ$C, the UV generation approaches saturation as we increase the pump power. This is caused by the photorefractive effect in lithium niobate, which is suppressed by increasing the temperature of the sample to 135$^\circ$C. At both the temperatures the pump wavelength was tuned so that the SFG output is phase matched at 369.525~nm. The external conversion efficiency is calculated from the pump power before the waveguide and the SFG power after the filters, while for the internal efficiency we accounted for pump coupling, losses from the optics and the Fresnel reflection chip facets of 17\% for the UV and 15\% for the pump. The IR coupled power was kept constant at 1~mW.

Figure \ref{FreqConv}b shows the DFG process at 135$^\circ$C obtained by replacing the IR laser with UV diode laser at 369.525~nm and a spectrum analyzer to detect the generated IR power, while the coupled UV power was kept constant at 30~$\mu$W.

The standard figure of merit to compare the quality of a waveguide design is the normalized efficiency of our device defined as:
\begin{equation}
\label{eq:eff_norm}
\eta_{norm}=		\frac{P_{out}}{P_{in}P_{pump}L^2}\textrm{,}
\end{equation}
where L is the interaction length of 4.9~mm and with values of 22.4~\%W$^{-1}$cm$^{-2}$ for SFG and 1.2~\%W$^{-1}$cm$^{-2}$ for DFG. While the single photon conversion efficiency is quite similar for SFG and DFG their values of  $\eta_{norm}$ are quite different. This is caused by the fact that during upconversion the energy per photon of the generated beam is increased while the opposite happens in downcoversion as quantified by the ratio of the wavelengths in Eq.~\ref{eq:eff}.

\section{Discussion on frequency conversion of single photons}\label{singleph}
\begin{figure}[b]
\centering
\includegraphics[width=0.45\textwidth]{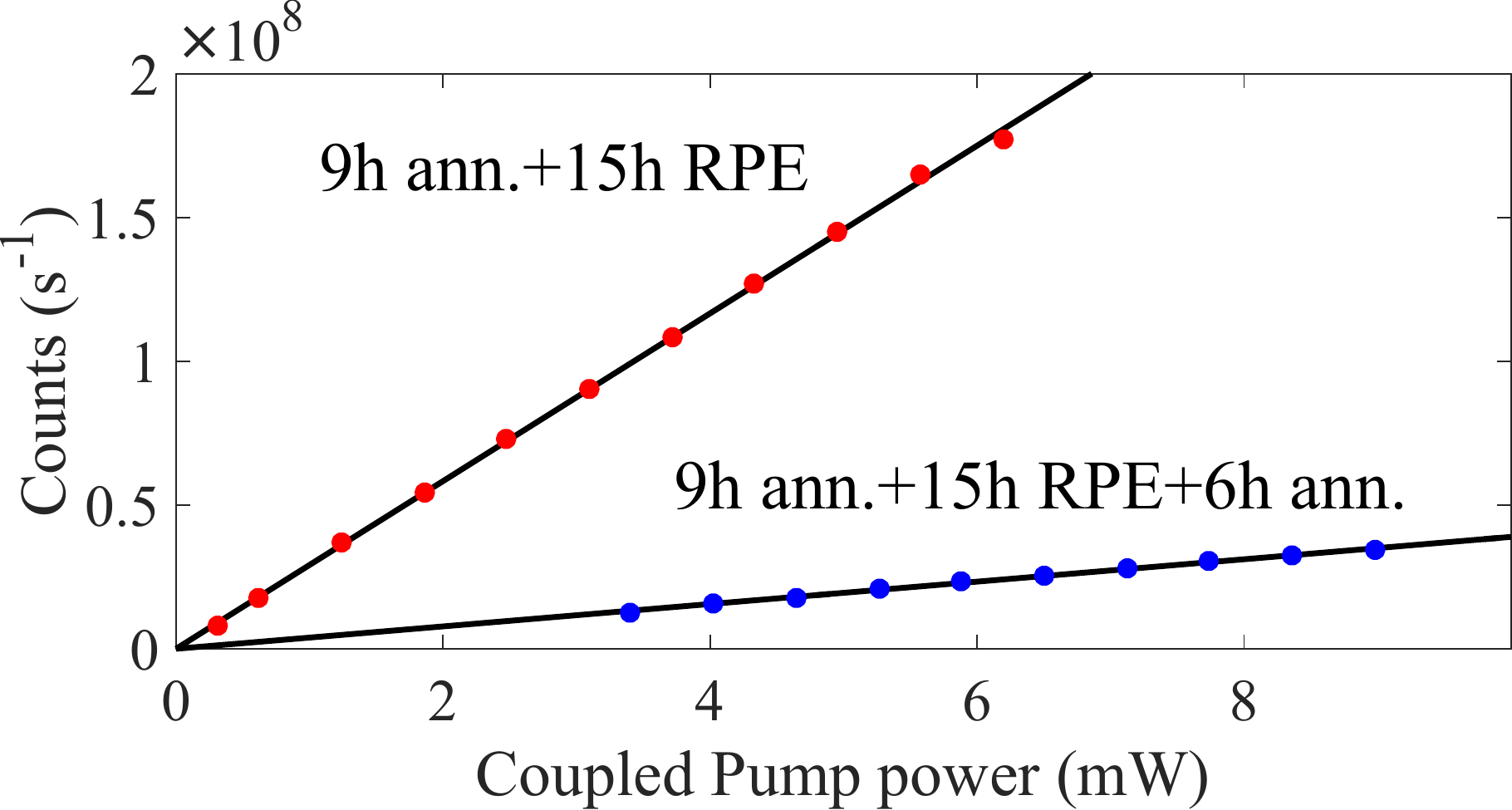}
\caption{Measure of the scattered Raman photons rate as a function of the coupled pump power after annealing and RPE (\textcolor[rgb]{1,0,0}{\textbullet}) and after 6h of extra annealing (\textcolor[rgb]{0,0,1}{\textbullet}) together with a linear fit (solid line). }
\label{noise}
\end{figure}
The conversion efficiency of our device was limited by its interaction length and pump power. However from the values of $\eta_{norm}$ in Eq.~\ref{eq:eff_norm} we can estimate a single photon conversion efficiency for a 3 cm interaction length and pump powers of 100~mW and 200~mW of 4.6\% and 9.2\% respectively for SFG and 4.5\% and 9.1\% for DFG. A further increase in efficiency could be achieved using first order quasi-phase-matching through innovative poling techniques that can reach sub-micron poling period \cite{BoesQPM}.

While the strong pump beam can be efficiently filtered out, it also produces spontaneously scattered Raman photons \cite{Pelc:11} that sit in the same wavelength range and affects the quality of the frequency conversion at the single photon level. Figure~\ref{noise} shows the measured scattered photons as a function of the pump power for the waveguide after annealing and RPE and 6h of extra annealing. After RPE the data are linear with a slope of 29$\times10^6$~counts/(s$\cdot$mW), this value is reduced by the extra annealing to 3.9$\times10^6$~counts/(s$\cdot$mW) because of the the mode of the pump being less confined. The Raman photons are spread over $\sim$20~nm around 1560~nm and less than 10$^5$~counts/s were measured from a 12~nm bandpass filter centered around 1570~nm for a coupled pump power of 5.9~mW pump. Since the light we are converting is resonant with a Yb$^+$ transition of 19.6~MHz and a lifetime of 8~ns, using a frequency filter of $\sim$100~MHz, readily available in the telecom regime, and a time gating of 20~ns we can reduce the probability of a noise photon in the time window below 4.6$\times10^{-6}$.

\section{Conclusions}\label{conclusions}
In conclusions we have demonstrated a nonlinear waveguide device capable of unifying trapped Yb$^+$ ions and standard telecom networks for quantum communication and quantum networking. We show UV and IR generation in the waveguide and measure the spontaneous Raman scattering generated by the pump laser. This interface is extremely versatile and can be used for the frequency conversion of time-bin and frequency encoded qubits. Finally, we assessed the potential performance of our technology for the conversion of single photons and the impact of the noise introduced by the waveguide. We estimate that attainable improvements in waveguide fabrication and pump power can achieve a conversion efficiency at the single photon level of 9\%. This efficiency could be further improved with a shorter poling period for first order quasi-phase-matching. We have also measured the rate of spontaneous Raman scattering for different waveguide configurations and its contribution to the single photon conversion process.
 
\ack{We thank Mojtaba Ghadimi, Valdis Bl\={u}ms and Benjamin G. Norton for their assistance with the UV laser and Elanor Huntington for lending us the single photon detector. This work has been supported by the Australian Research Council (ARC) under the Grants DP140100808 and Griffith University Research Infrastructure Program. ML acknowledges the support of the ARC-Decra DE130100304. EWS was supported by ARC Future Fellowship FT130100472. A.M. acknowledge the support of the Australian Research Council (ARC) Centre of Excellence Funding (CE110001018). This work was performed in part at the Queensland node - Griffith - of the Australian National Fabrication Facility, a company established under the National Collaborative Research Infrastructure Strategy to provide nano and microfabrication facilities for Australia's researchers.} 

\vspace{1pc}
\bibliography{Kasture-bib-file}{}
\bibliographystyle{iopart-num}
\end{document}